\documentclass[11pt]{article}
\usepackage{moriond,epsfig}

\def\be{\begin{equation}}
\def\ee{\end{equation}}
\def\bea{\begin{eqnarray}}
\def\eea{\end{eqnarray}}

\begin{document}
\vspace*{2cm}

\hfill{UCHEP-10-02}

\title{STRANGE BEAUTY AND OTHER BEASTS FROM $\Upsilon$(5S) AT BELLE
\footnote{talk presented at the XLV$^{th}$ Rencontres de Moriond, 6-13 March, 2010}
}

\author{K. Kinoshita}

\address{University of Cincinnati,
PO Box 210011 \\ 
Cincinnati, OH 45221, USA 
}

\maketitle\abstracts{
The $B$-factories have successfully exploited the unique advantages of the $\Upsilon$(4S) resonance to study many aspects of  $B_d$ and $B_u$ mesons.
The $\Upsilon$(10860) (aka $\Upsilon$(5S)) resonance, which is above mass threshold for the $B_s$ and shares many of the same advantages, has been relatively unexplored.  
The Belle experiment has collected more than 120 fb$^{-1}$ at the $\Upsilon$(10860) and 7.9 fb$^{-1}$ at higher energies, corresponding to more than 7 million $B_s$ events.  
Recent results based on $\approx$20\% of these data are presented and prospects for future possiblities discussed. 
}


The Belle experiment,\cite{belle} located at KEKB,\cite{kekb}  was built primarily to measure  $CP$ asymmetries of $B$ meson decay in $e^+e^-$ annihilations at the $\Upsilon$(4S) resonance.
In December 2009, the integrated luminosity surpassed 1000~fb$^{-1}$ (=1~ab$^{-1}$), fulfilling the goal stated in the original Belle proposal.
Not all of the data were taken at the $\Upsilon$(4S), however -- results from the $\Upsilon$(10860) resonance are presented here.

The $\Upsilon$(10860), ($Mc^2=10865\pm 8$~MeV/$c^2$,  $\Gamma=110\pm 13$~MeV),\cite{pdg} is interpreted as $\Upsilon$(5S), the fourth excitation of the vector bound state of $b\bar b$.
It is above $B_s\bar B_s$ threshold and, given the success of the $\Upsilon$(4S) program in characterizing properties of  $B_{d,u}$, it is natural to contemplate $B_s$ at $\Upsilon$(10860).
The $e^+e^-$ environment produces clean events, efficiently triggered,  with precisely known center-of-mass energy.
Furthermore, the $B$-factory offers an existing facility with high luminosity, a well-studied detector with precise and sensitive photon detection, and an abundance of $\Upsilon$(4S) data for comparisons.
While rates are low compared to hadronic collisions,
$\sigma(e^+e^-\to \Upsilon$(10860))$\approx \sigma(e^+e^-\to \Upsilon$(4S)$)/3\approx 0.3$~nb, and events include $B_{d,u}$ as well as $B_s$, the $\Upsilon$(10860) can be competitive nonetheless, particularly in aspects of $B_s$ decay that are limited by systematic effects at a hadron machine.

Belle has collected data at the $\Upsilon$(10860) in several runs.  
In June 2005 a three-day ``engineering'' run served to test KEKB, which had never operated at energies above the $\Upsilon$(4S), and study the basics of $\Upsilon$(10860), $B_s$, and $B_s^*$.
A scan of five energy points was used to locate the peak, at $\sqrt{s}=$10869~MeV, where 1.86~fb$^{-1}$ were collected.
By June 2006, results confirmed the projected potential of $\Upsilon$(10860), and 21.7~fb$^{-1}$ were collected in 20 days.
In December 2007, $\approx 8$~fb$^{-1}$ were collected in a scan of six energy points near and above the $\Upsilon$(10860).
Finally, a large fraction of the data collected in October 2008-December 2009 have been at the resonance.
The integrated luminosity collected on resonance over all data sets thus far is $\approx 120$~fb$^{-1}$.

\begin{figure}[t]
\begin{center}
{\epsfysize=4.3cm\epsfbox{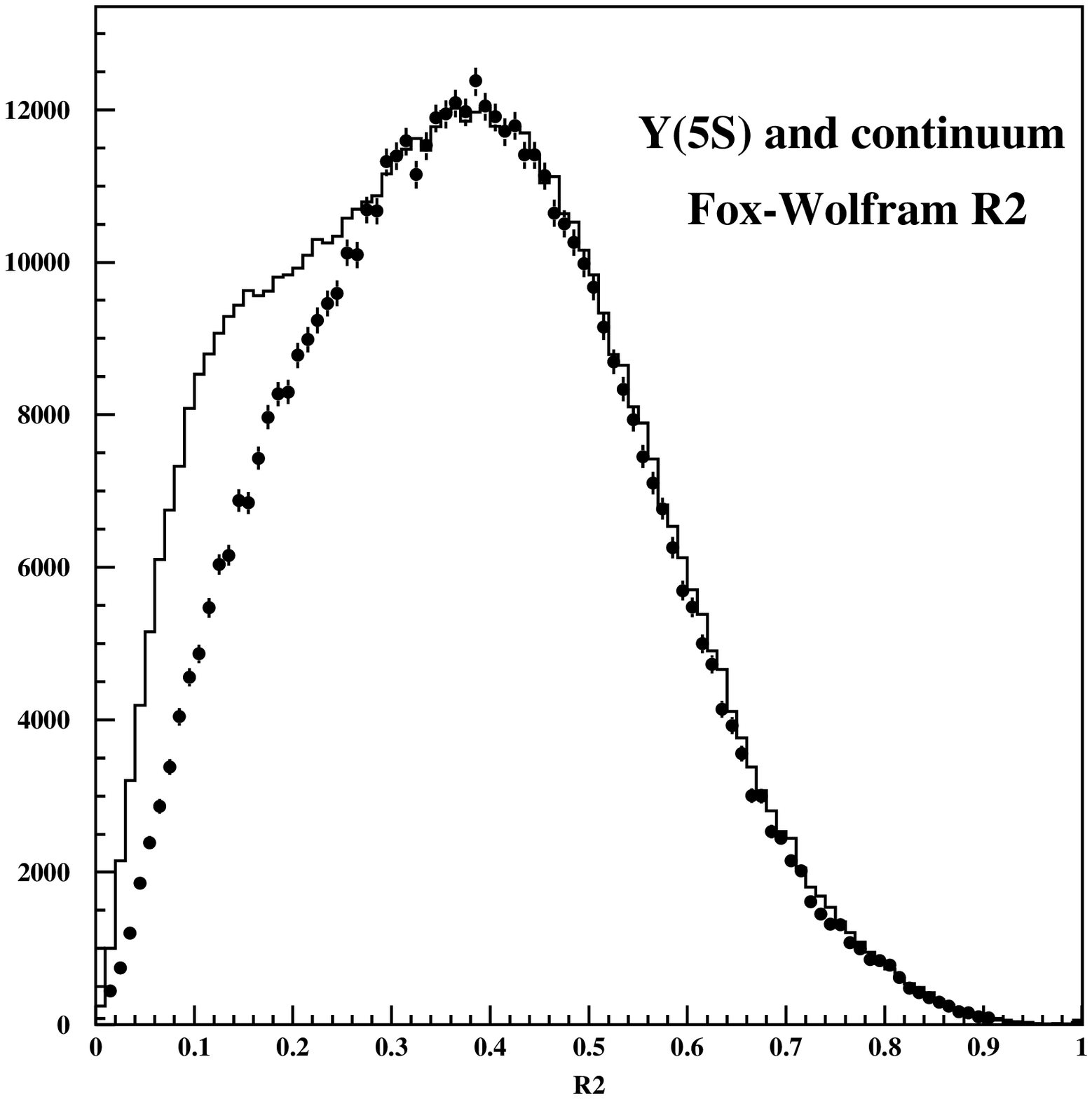}}~{\epsfysize=4.5cm\epsfbox{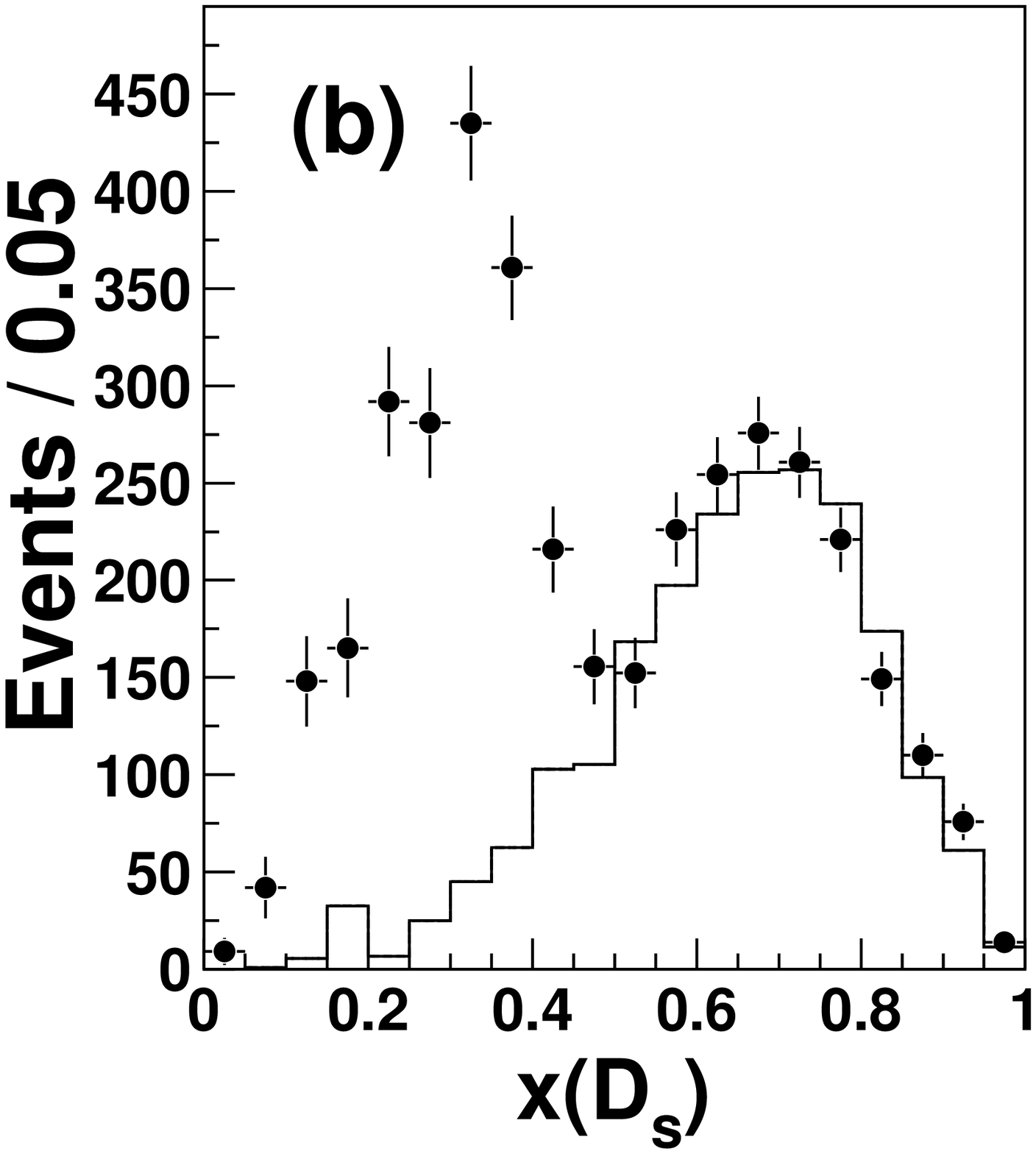}} 
\end{center} 
\caption{(L) Distribution in $R_2$, (histogram) 1.86~fb$^{-1}$ at the $\Upsilon$(10860) and  (points) continuum below  $\Upsilon$(4S), scaled. (R) Distribution of $D_s$ in $x\equiv p_{D_s}/\sqrt{E_{beam}^2-M_{D_s}^2}$, (points) $\Upsilon$(10860) and (histogram) scaled  continuum.
 \label{5S_profile}}
\end{figure}

As prerequisite to studies of $B_s$, its abundance in $\Upsilon$(10860) events was determined from the 2005 data.\cite{5Sinclusive}
About 10\% of the hadronic events are resonance (assumed to be $b\bar b$), the rest being continuum $e^+e^-\to q\bar q$ ($q=u,d,s,c$).
Figure~\ref{5S_profile} (L) displays $R_2$, the ratio of the 2$^{nd}$ and 0$^{th}$ Fox-Wolfram moments,\cite{foxwolfram} a measure of ``jettiness'' that tends to be lower for the more isotropic resonance events.
We find $(3.01\pm0.02\pm 0.16)\times 10^5$~$b\bar b$~events/fb$^{-1}$.
The $b\bar b$ events may fragment to:
$B_s^{(*)} \bar B_s^{(*)}$, $B_q^{(*)} \bar B_q^{(*)}$, $B_q \bar B_q^{(*)}\pi$, $B_q \bar B_q\pi\pi$ ($q$ is a $u$- or $d$-quark).
The fraction ($f_s$) that are $B_s^{(*)} \bar B_s^{(*)}$ is determined through measurement of the inclusive rate ${\mathcal B}(\Upsilon(10860)\to D_s X)\equiv{\mathcal B}_{\Upsilon}$  (Figure~\ref{5S_profile}(R)), an average over $B_s$, $B_d$, and $B_u$ weighted by abundance, combined with the rate for $B_{u,d}\to D_s X$ measured at the $\Upsilon$(4S) and a semi-theoretical estimate ${\mathcal B}(B_s\to D_s X)=(92\pm 11)\%$,\cite{cleo_5S}
under an assumption that $B_d$ and $B_u$ are produced equally and that non-$B$ production is negligible.
The inclusive $D^0$ rate gives an independent value of $f_s$ with larger uncertainties;  ${\mathcal B}(B_s\to D^0X)\ll {\mathcal B}(B_q\to D^0X)$.
The results are combined to obtain $f_s=(18.0\pm 1.3\pm 3.2)\%$.\cite{5Sinclusive}

\begin{figure}[ht]
\begin{center}
{\epsfxsize=5.5cm\epsfbox{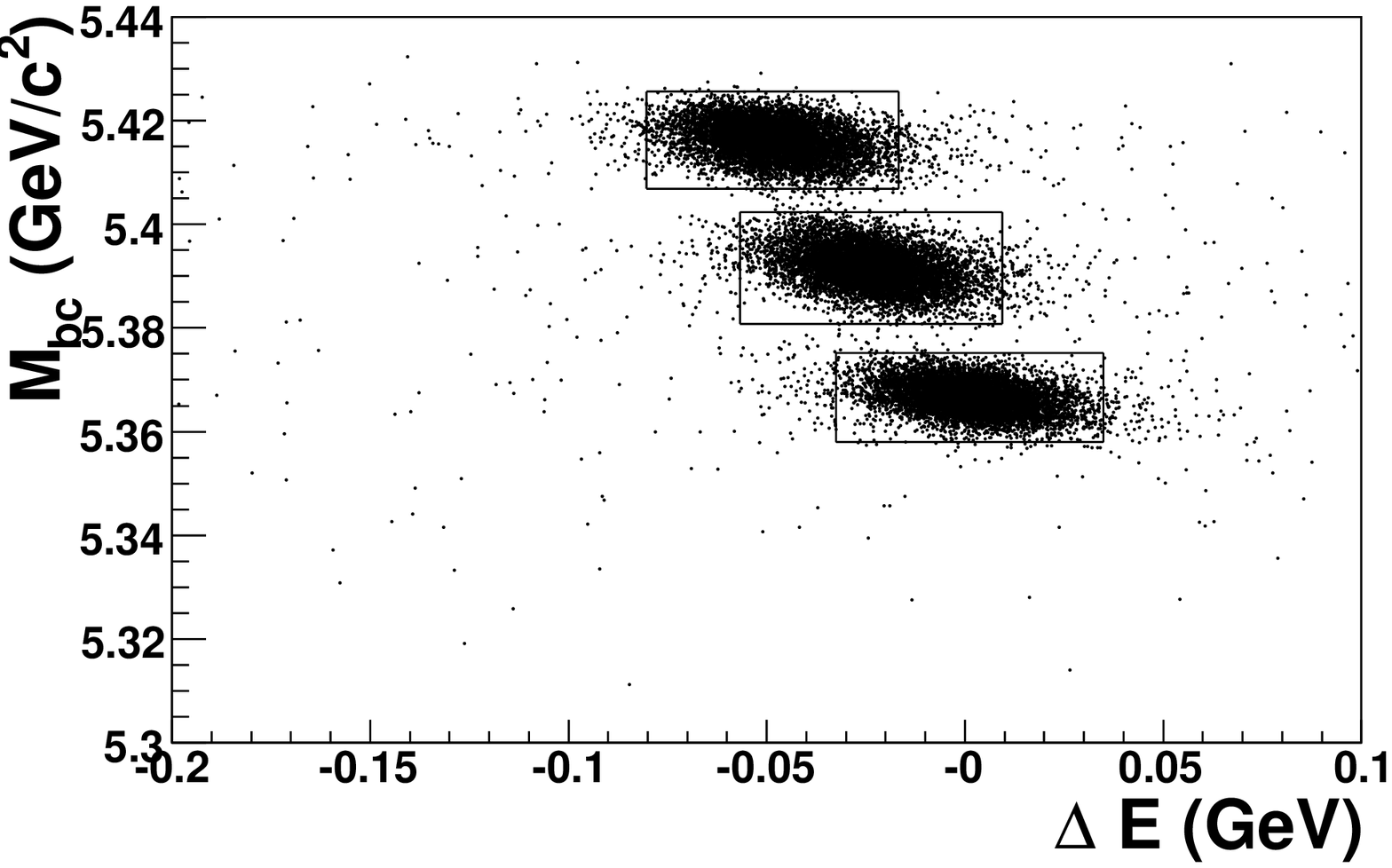}}{\epsfxsize=5.5cm\epsfbox{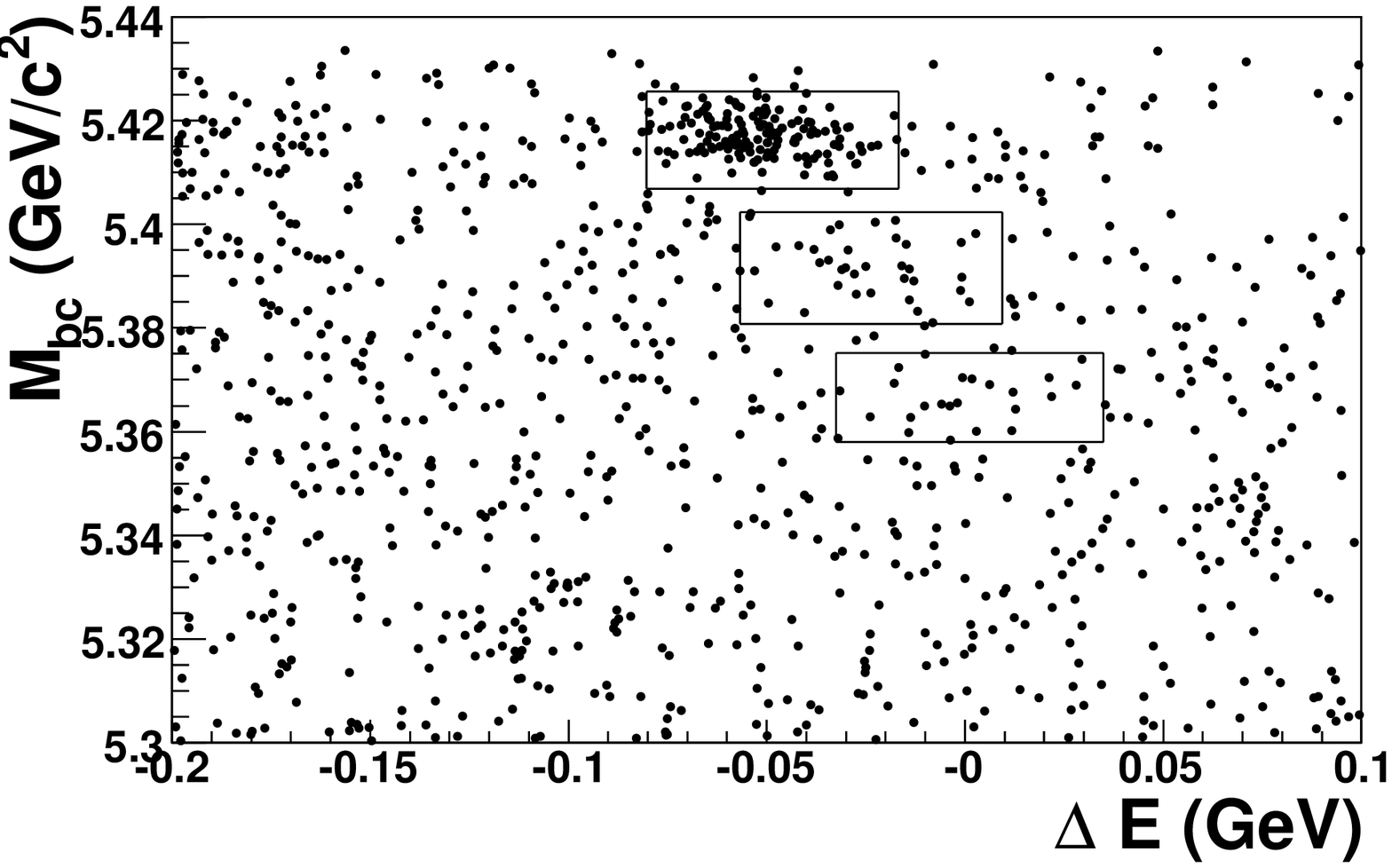}} 
\end{center} 
\caption{Illustration of full reconstruction  method,  $B_s\to D_s\pi$.
Distributions in $\Delta$E and $M_{\rm bc}$ of candidates, (Left) Monte Carlo simulation, (Right) data, 23.6~fb$^{-1}$.
Also shown are  signal regions for  $B_s^*\bar B_s^*$(upper signal box), $B_s^*\bar B_s$(middle box), and $B_s\bar B_s$(lower box) events.
\label{fig:Dspi}}
\end{figure}

At the energy of the $\Upsilon$(10860), three types of $B_s$ events are allowed: $B_s$ pairs, $B_s^*$ pairs, and mixed $B_s\bar B_s^*$ events.
These three types may be separated through ``full reconstruction'' of $B_s$ decays, where all decay products are measured, a method used with great success for $B_q$ at the $\Upsilon$(4S).
Each candidate's energy and momentum in the $e^+e^-$ center-of-mass are evaluated as $\Delta E \equiv E_{cand}-E_{beam}$ and $M_{\rm bc}\equiv \sqrt{E_{beam}^2-p_{cand}^2}$.
In  $B_s\bar B_s$ events (analogous to $\Upsilon$(4S)$\to B_q\bar B_q$), the $B_s$ carries the beam energy, so  $\left< \Delta E\right> =0$ and $\left< M_{\rm bc}\right> =M_{B_s}$.
For $B_s^*\bar B_s$ or $B_s^*\bar B_s^*$, the kinematics of $B_s^*\to B_s\gamma$ ($E_\gamma=50~{\rm MeV}$) leads to localized signals as well.
In the case of $B_s^*\bar B_s$ the end result is effectively the loss of $50$~MeV from the $B_s$ pair, with the energy difference being shared approximately equally: $\left< \Delta E\right> \approx -25~$MeV and $\left< M_{\rm bc}\right>\approx M_{B_s}+25$~MeV.
For $B_s^*\bar B_s^*$ the corresponding energy reduction is $\sim$50~MeV.
Figure~\ref{fig:Dspi} shows distributions in $\Delta E$ and $M_{\rm bc}$ for $B_s\to D_s^-\pi^+$ candidates, signal Monte Carlo simulations and (previously reported) data.\cite{Dspi}  
The $B_s$ yields are extracted from a two-dimensional fit in these two variables, unless otherwise specified.
From the relative yields of the three modes, we find \cite{Dspi} 
\begin{eqnarray*}
f_{B_s^*B_s^*}
&\equiv& {\sigma(e^+e^-\to B_s^*\bar B_s^*)\over \sigma(e^+e^-\to B_s^{(*)}\bar B_s^{(*)})}=(90.1^{+3.8}_{-4.0}\pm 0.2)\% \\
f_{B_s^*B_s}&\equiv& {\sigma(e^+e^-\to B_s^*\bar B_s + B_s\bar B_s^*))\over \sigma(e^+e^-\to B_s^{(*)}\bar B_s^{(*)})} =(7.3\pm 0.3\pm 0.1)\% 
\end{eqnarray*}


The new results presented here, based on 23.6~fb$^{-1}$ of data collected in 2005-6,  include $B_s\to D_s^{*-}\pi^+$, $D_s^{(*)-}\rho^+$, $B_s\to D_s^{(*)+}D_s^{(*)-}$, $B_s\to J/\psi \eta^{(\prime)}$, $B_s\to hh$, and $\Upsilon$(5S)$\to B\bar BX$.

\begin{figure}[t]
\begin{center}
{\epsfxsize=5.2cm\epsfbox{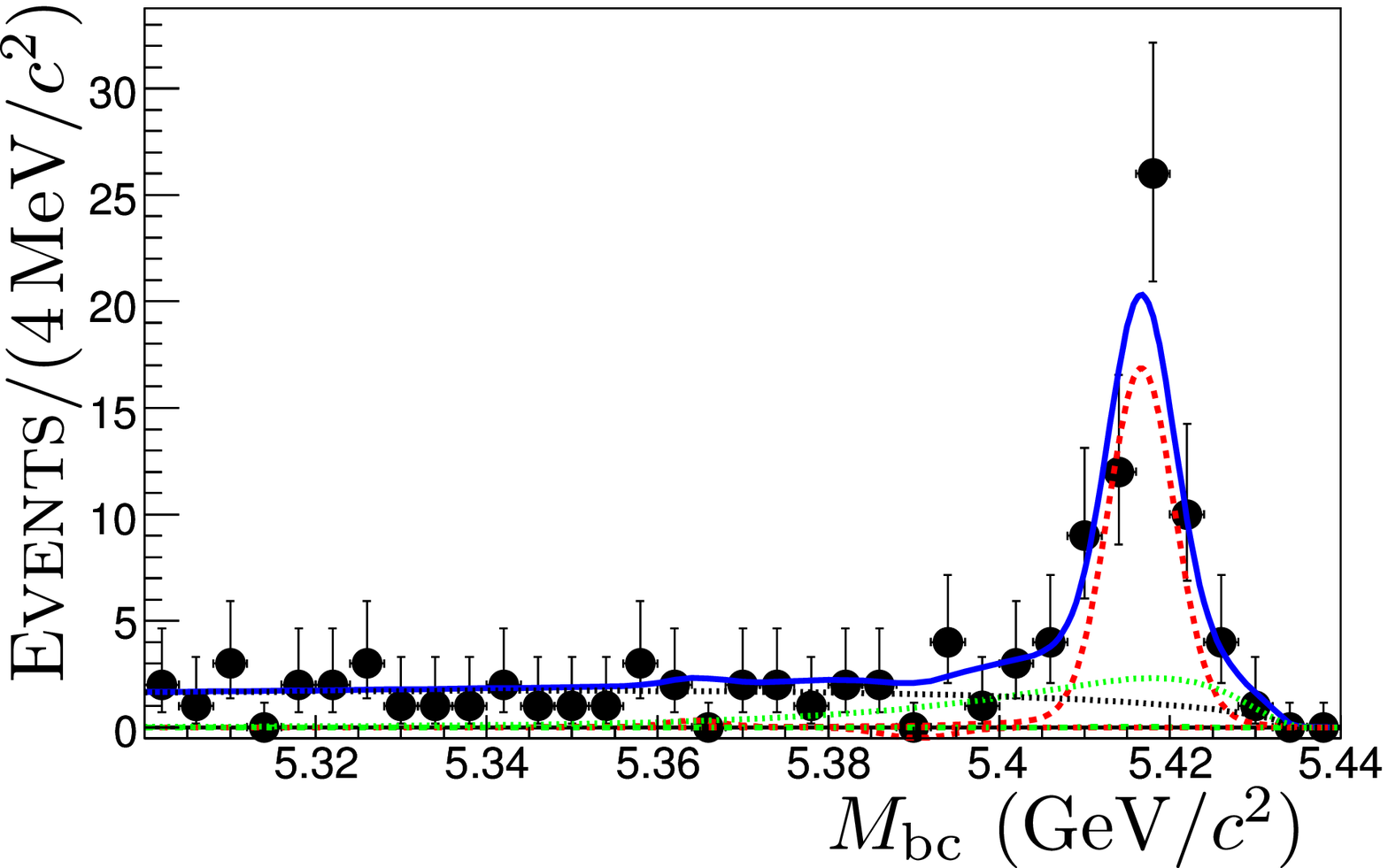}}{\epsfxsize=5.2cm\epsfbox{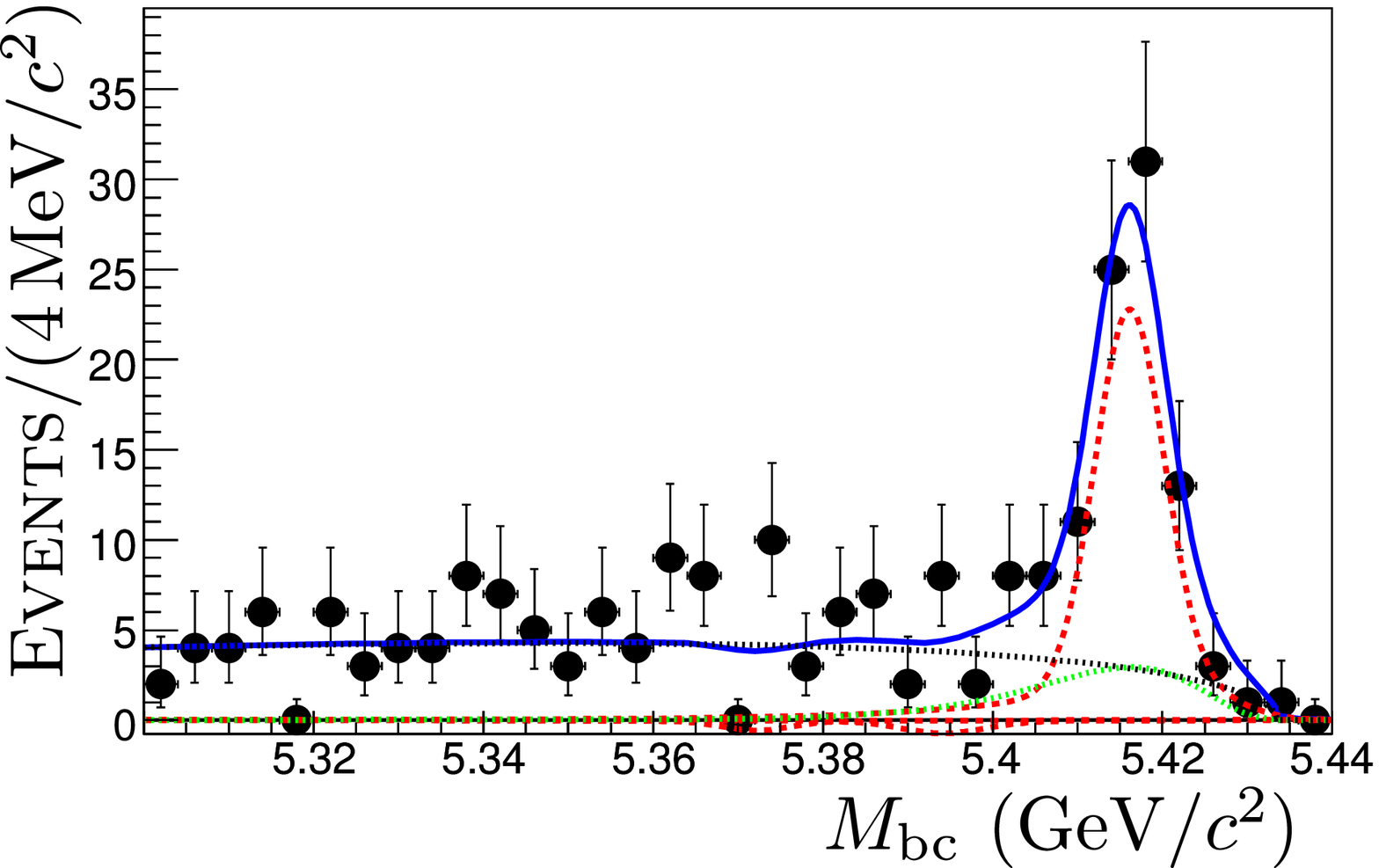}} {\epsfxsize=5.2cm\epsfbox{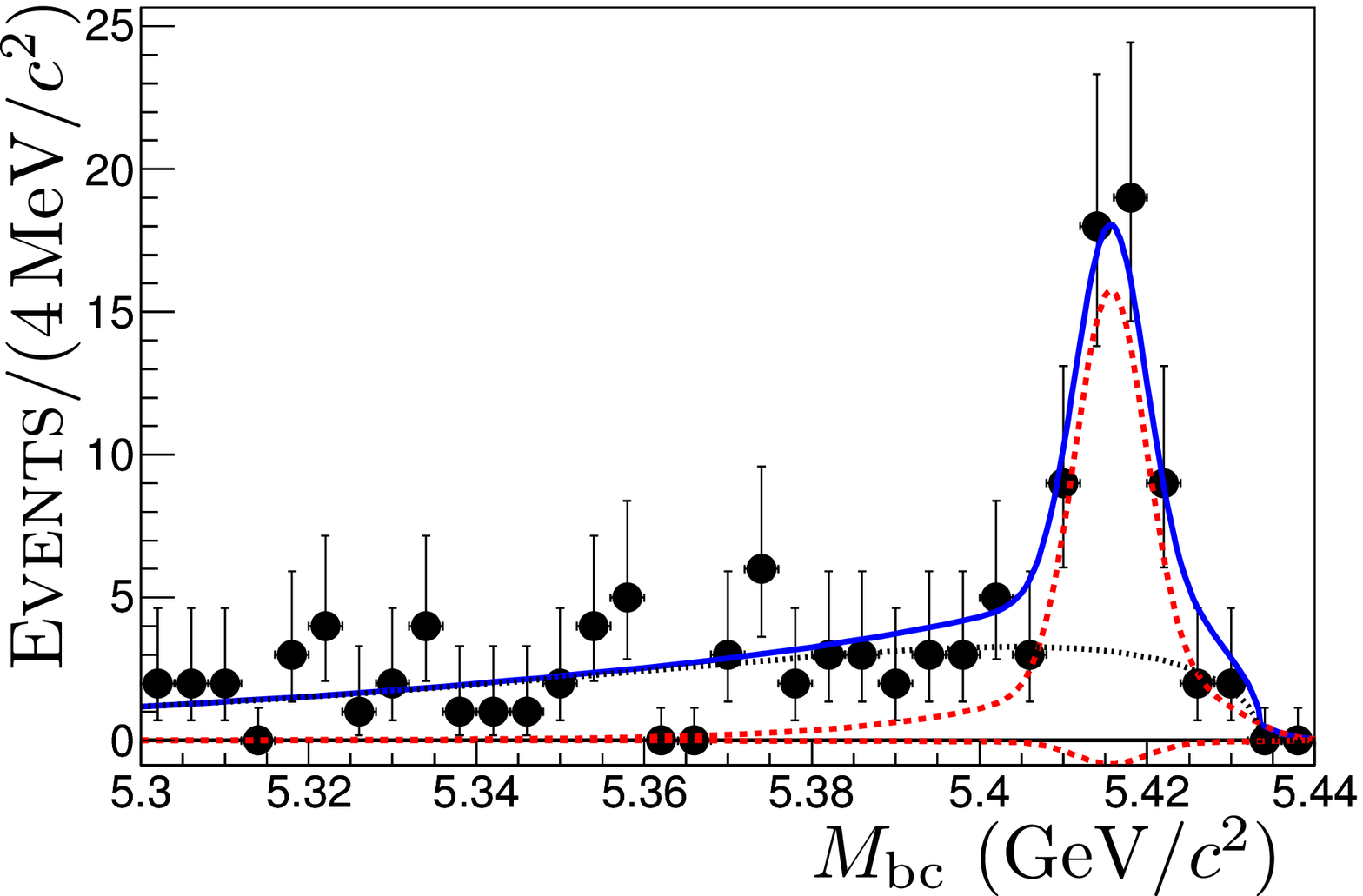}} 
\end{center} 
\caption{Projections in  $M_{\rm bc}$ for 23.6~fb$^{-1}$ data:  (left) $B_s\to D_s^*\pi$,  (center) $B_s\to D_s\rho$, (right) $B_s\to D_s^*\rho$.
\label{Dssth}}
\end{figure}

\begin{figure}[t]
\begin{center}
{\epsfxsize=5.5cm\epsfbox{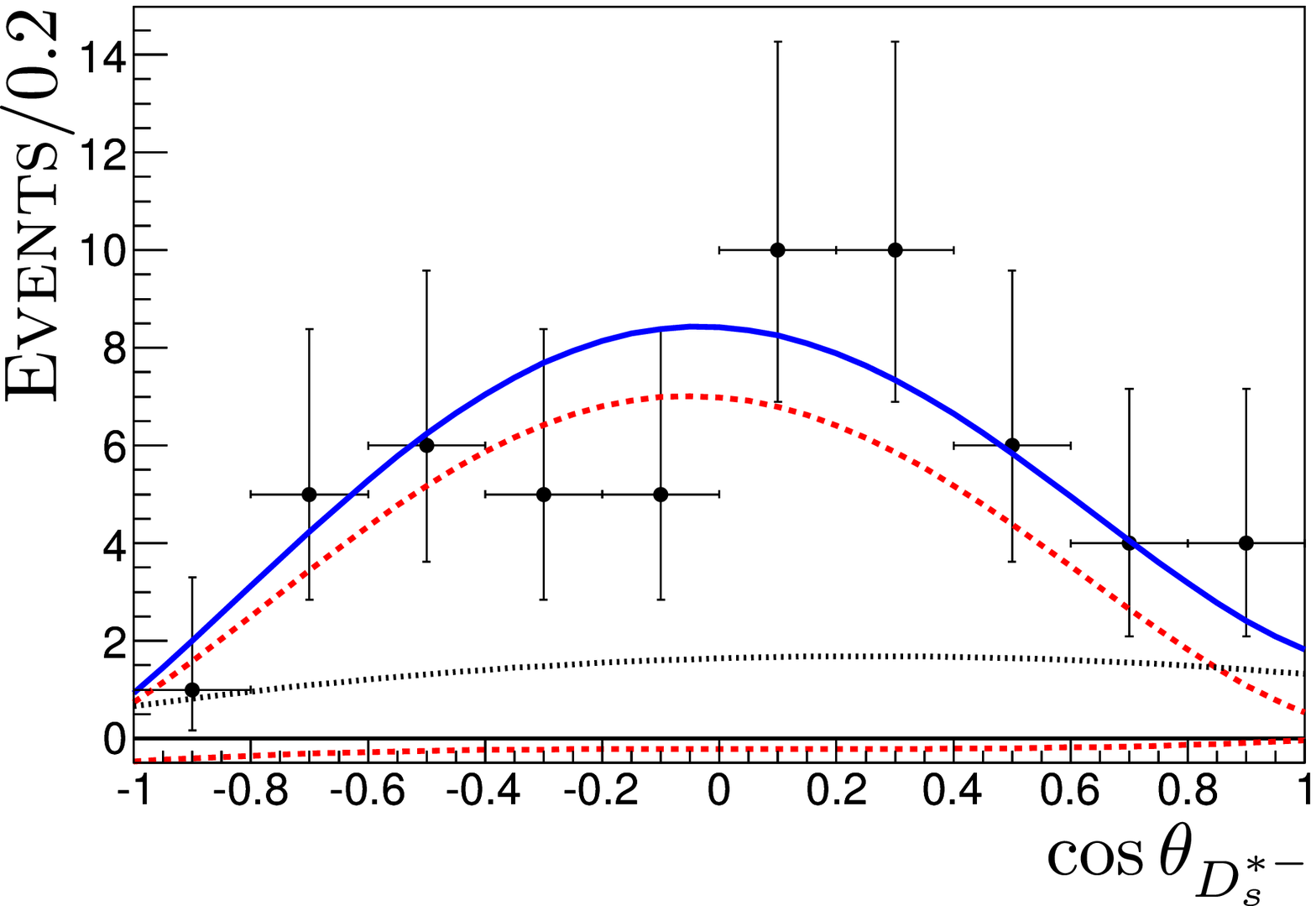}}{\epsfxsize=5.5cm\epsfbox{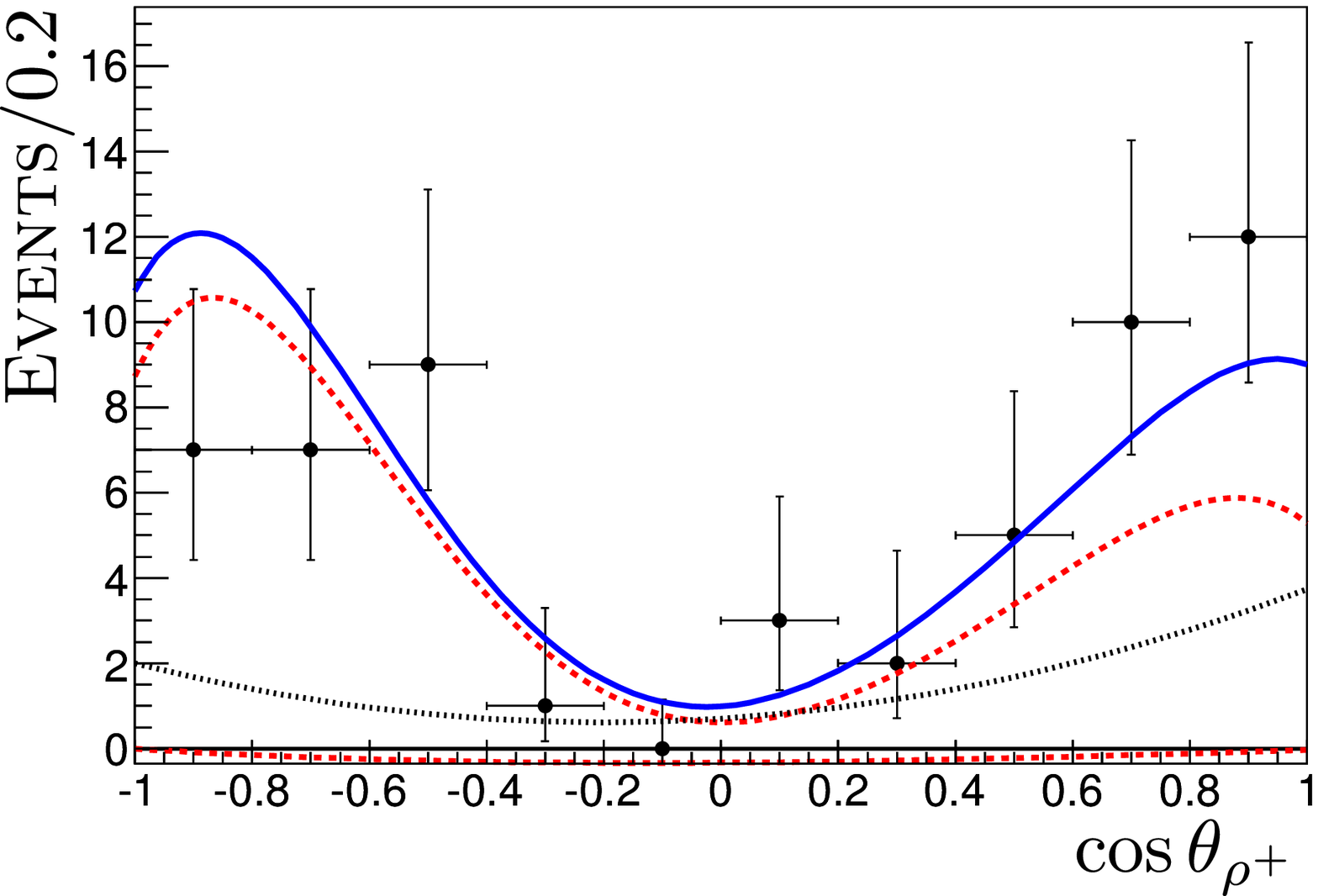}} 
\end{center} 
\caption
{Projections in helicity angles of  $B_s\to D_s^*\rho$ signal events: (left) $\cos_{D_s^*}$, (right) $\cos_\rho$.
}
\label{Dsstangle}
\end{figure}

The decays $B_s\to D_s^{(*)-}h^+$, where $h$ is a light non-strange meson, proceed dominantly via a CKM-favored spectator process.
We reconstruct $D_s$ in the modes
$\phi (\to K^+K^-)\pi^-$, $K^{*0}(\to K^+K^-)K^-$, and $K_S(\to \pi^+\pi^-)K^-$. 
Shown in Figure~\ref{Dssth} are fits to data, projected into $M_{\rm bc}$, for  $B_s\to D_s^{*-}\pi^+$, $D_s^{-}\rho^+$, and $D_s^{*-}\rho^+$.
Two-dimensional fitting for $B_s^*\bar B_s^*$ only yields signals  $53.4^{+10.3}_{-9.4}$ ($7.1\sigma$) and $92.2^{+14.2}_{-13.2}$ ($8.2\sigma$)  in  $B_s\to D_s^{*-}\pi^+$ and $B_s\to D_s^{-}\rho^+$, respectively.
Taking $f_{B_s^*B_s^*}=90.1\%$, $f_s=(19.5^{+3.0}_{-2.3})\%$, and $\sigma_{e^+e^-\to b\bar b}=0.302\pm 0.014$~nb (a weighted average from \cite{5Sinclusive,cleo_5Sb}), we find
${\cal B}=(2.4^{+0.5}_{-0.4}\pm 0.3\pm 0.4)\times 10^{-3}$ and ${\cal B}=(8.5^{+1.3}_{-1.2}\pm 1.1\pm 1.3)\times 10^{-3}$.\cite{Dssth}
For each result, first error is statistical, the third is systematic (due to the uncertainty in $f_s$), and the second is systematic (from all other uncertainties).
For  $B_s\to D_s^{*-}\rho^+$, a pseudoscalar decay to two vectors, the distributions in the helicity angles $\theta_{D_s^{*-}}$ and $\theta_{\rho^+}$ depend on the relative contribution from the different helicity states, which depends on the detailed hadronization mechanism for the decay; for example, the factorization hypothesis predicts that longitudinal polarization dominates: $f_L\approx 88\%$.\cite{factorization}
A four-dimensional fit yields $77.7^{+14.6}_{-13.3}$ ($7.4\sigma$) signal events
from which we derive a branching fraction $(11.8^{+2.2}_{-2.0}\pm 1.7\pm 1.8)\times 10^{-3}$.\cite {Dssth}
We also measure $f_L = 1.05{^{+0.08}_{-0.10}} {^{+0.03}_{-0.04}}$.\cite{Dssth}


\begin{figure}[ht]
\begin{center}
{\epsfxsize=5.2cm\epsfbox{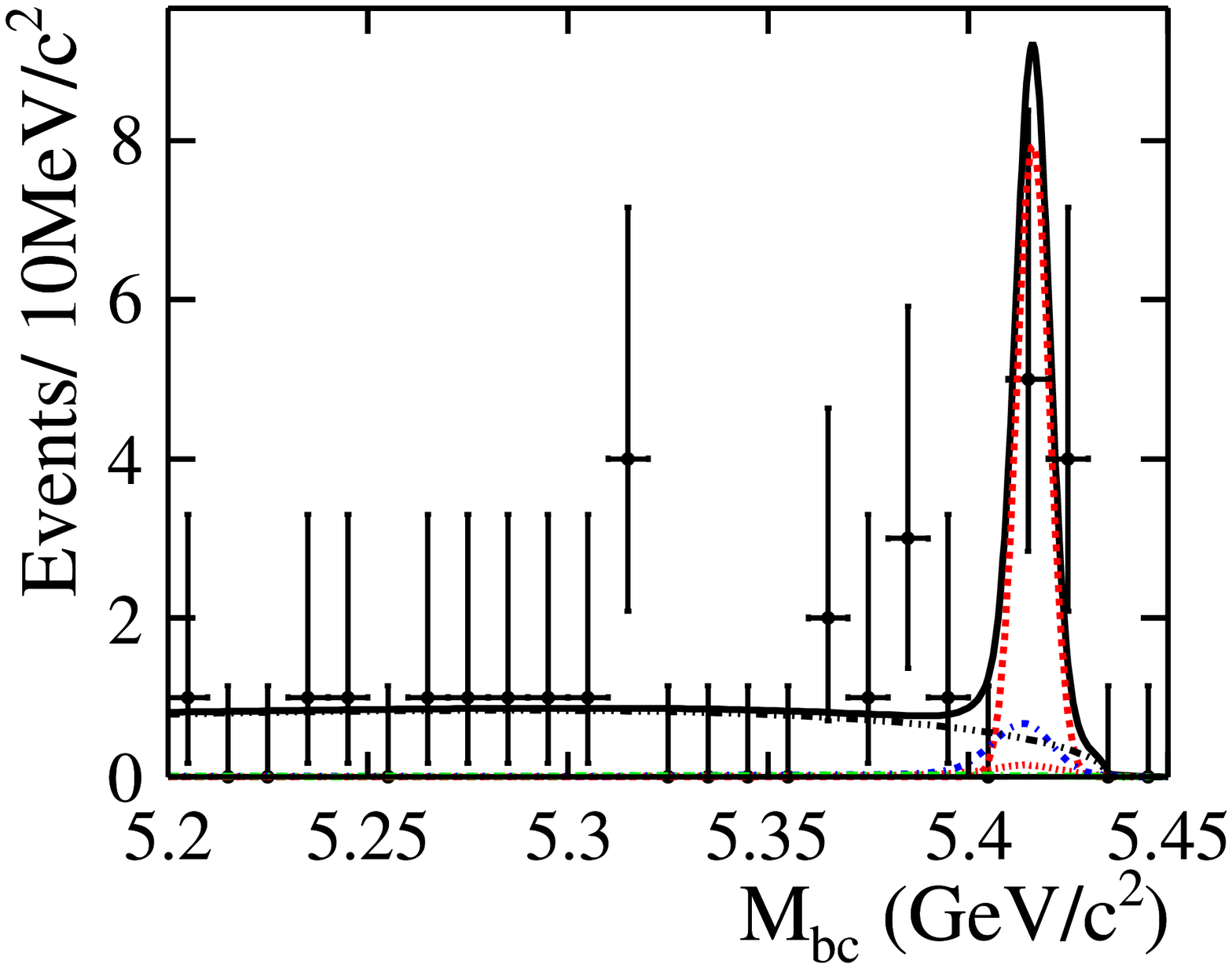}}{\epsfxsize=5.2cm\epsfbox{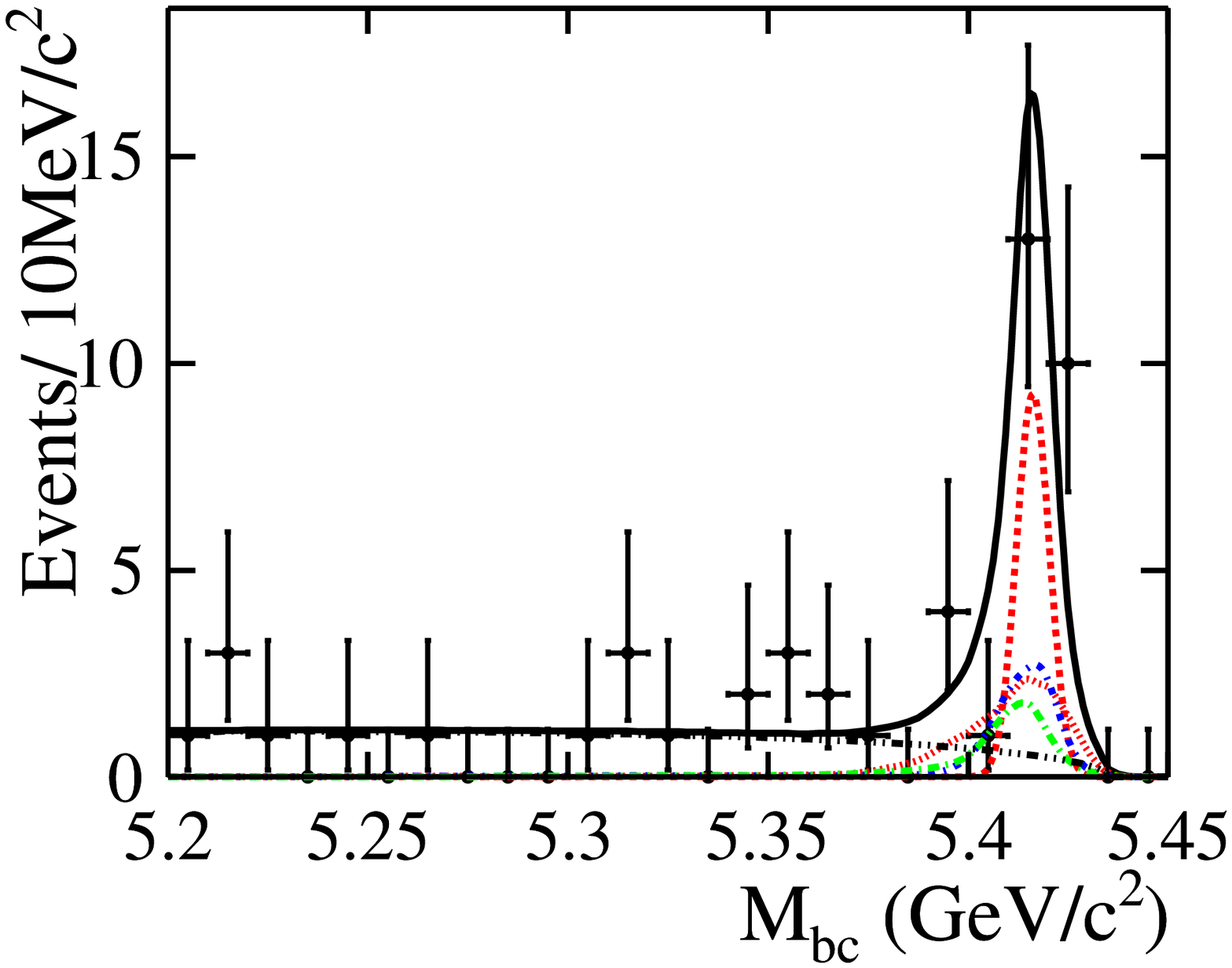}} {\epsfxsize=5.2cm\epsfbox{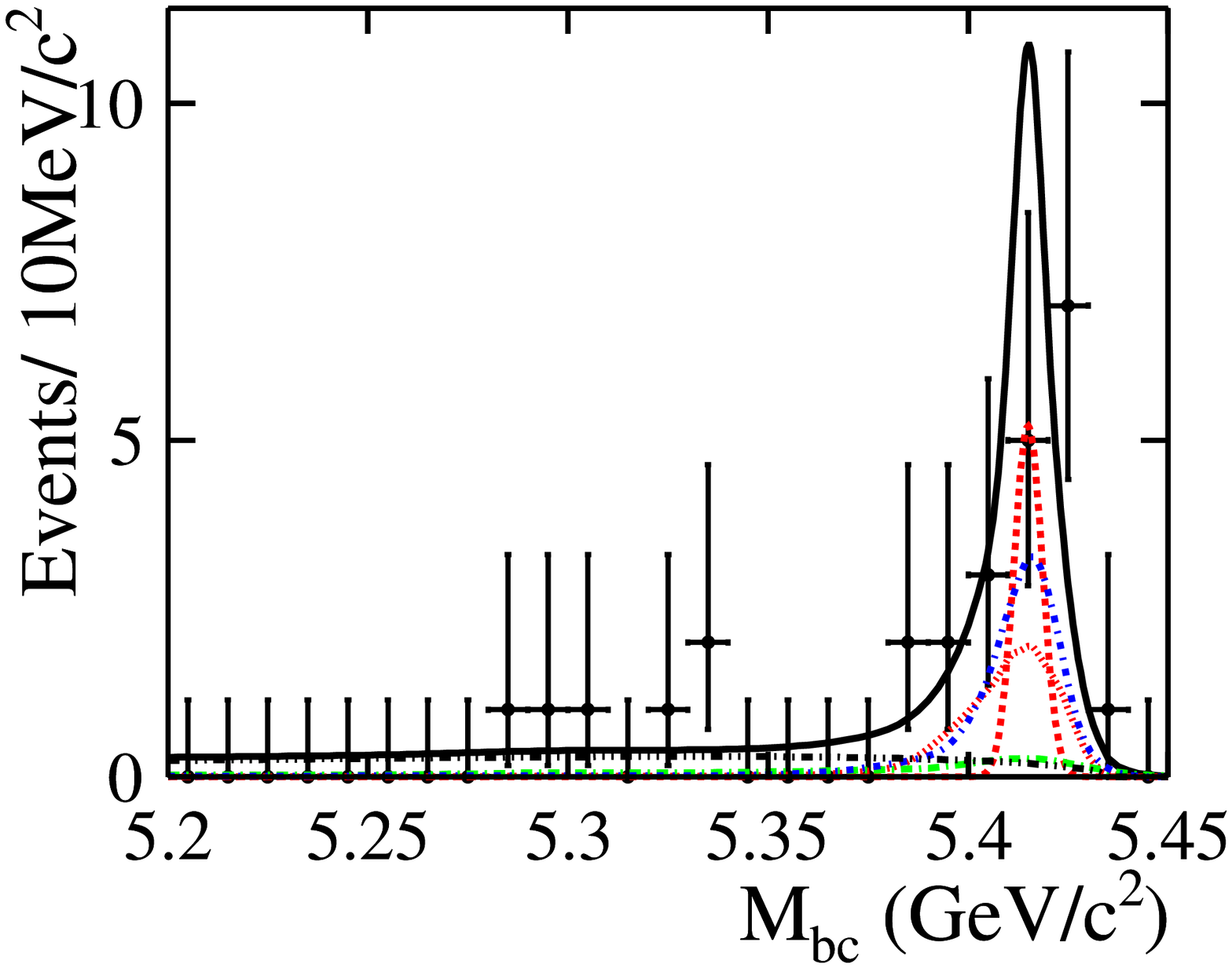}} 
\end{center} 
\caption{Projections in  $M_{\rm bc}$, based on 23.6~fb$^{-1}$ of Belle data at $\Upsilon$(10860):  (left) $B_s\to D_s^+D_s^-$,  (center) $B_s\to D_s^{*+}D_s^-$, (right) $B_s\to D_s^{*+}D_s^{*-}$ curves show the fitted total (blue), signal (dashed), wrong combination (dotted; red), cross-feed (dash-dotted; blue), combinatorial (dash-3dot).
\label{DsDs}}
\end{figure}

If the $B_s$  decays via the spectator process where the $W^-$ couples to $s\bar c$, the net flavor of the final state is zero: $c\bar s \bar c s$.
If the $b$- and $c$-quark masses are infinite, $CP$=+1.
Because the state is not far above mass threshold, two-body final states, $B_s\to D_s^{(*)-}D_s^{(*)+}$, are expected to dominate.\cite{DsDs_Dunietz}
As the $s\bar c$ coupling is CKM-favored, this channel comprises a substantial fraction of $B_s$ decay channels, of order 10\%, resulting in a substantial asymmetry in lifetime between the two $CP$ eigenstates that is related to the branching fraction for $B_s\to D_s^{(*)-}D_s^{(*)+}$:
${\Delta \Gamma_{CP}\over \Gamma}\approx {2{\cal B} (B_s\to D_s^{(*)+} D_s^{(*)-})\over 1- {\cal B} (B_s\to D_s^{(*)+} D_s^{(*)-})}$.\cite{aleksan}
The quantity of interest here is the sum, which is difficult to measure at hadron machines due to low efficiencies for detection of the photon from $D_s^*\to D_s\gamma$. 

We reconstruct $B_s\to D_s^{(*)-}D_s^{(*)+}$ candidates in the following modes:
$D_s^{*+}\to D_s^{+}\gamma$; $D_s^+\to\phi\pi^+$, $K_S^0K^+$, $\bar K^{*0}K^+$, $\phi\rho^+$, $K^{*+} K_S^0$, $K^{*+}\bar K^{*0}$; $\phi\to K^+K^-$; $K_S^0\to \pi^+\pi^-$; $\bar K^{*0}\to K^-\pi^+$; $\rho^+\to \pi^+\pi^0$; $K^{*+}\to K_S^0\pi^+$.
Candidates are pre-selected by requiring  $5.2<M_{\rm bc}c^2/{\rm GeV}<5.45$ and $-0.15<\Delta E/{\rm GeV}<0.1$.
In each event the candidate with the lowest $\chi^2$ based on $M(D_s)$ and $M(D_s^*)-M(D_s)$ is selected.
The yield is evaluated through a simultaneous fit, over all three modes $B_s\to D_s^{(*)-}D_s^{(*)+}$, accounting for signal, combinatorial background, crossfeed from the other signal modes, and wrong combinations in signal events.
Projections onto $M_{bc}$ are shown in Figure~\ref{DsDs} and the yields and branching fractions in Table~\ref{table:DsDs}.
Taking the sum, ${\cal B}=6.9{^{+1.5}_{-1.3}\pm 1.9}$, we calculate ${\Delta\Gamma_{CP}\over \Gamma} = 0.147^{+0.036}_{-0.030}{^{+0.044}_{-0.042}}\pm 0.004$, where the last error is the estimated theory error.
This value is consistent with the current PDG value, $0.092^{+0.051}_{-0.054}$.

\begin{table}[t]
\caption{Yields, branching fractions, and signal significance for $B_s\to D_s^{(*)-}D_s^{(*)+}$.\label{table:DsDs}}
\vspace{0.4cm}
\begin{center}
\begin{tabular}{|c c c c|}
\hline
Mode & Yield & ${\cal B}$(\%) & $S$ \\
\hline
$D_s^{-}D_s^{+}$& $8.5{^{+3.2}_{-2.6}}$& $1.0{^{+0.4}_{-0.3}{^{+0.3}_{-0.2}}}$ & 6.2 \\
$D_s^{-}D_s^{*+}$& $9.2{^{+2.8}_{-2.4}}$ & $2.8{^{+0.8}_{-0.7}\pm 0.7}$ & 6.6\\
$D_s^{*-}D_s^{*+}$& $4.9{^{+1.9}_{-1.7}}$& $3.1{^{+1.2}_{-1.0}\pm 0.8}$ & 3.2 \\
\hline
\end{tabular}
\end{center}
\end{table}

In the Standard Model, mixing-mediated $CP$-violation occurs in neutral mesons due to the complex argument of the product of CKM matrix elements of the mixing ``box diagram.'' 
For $B_s$ the relevant product is $V_{\rm tb}^{*2}V_{\rm ts}^2$, which is real, so that no significant asymmetry is expected.
$CP$-asymmetries in decays of $B_s$ thus present an opportunity to reveal New Physics.
These measurements will require the reconstruction of a sizable sample of $CP$-defined final states.  

\begin{figure}[ht]
\begin{center}
{\epsfxsize=5.2cm\epsfbox{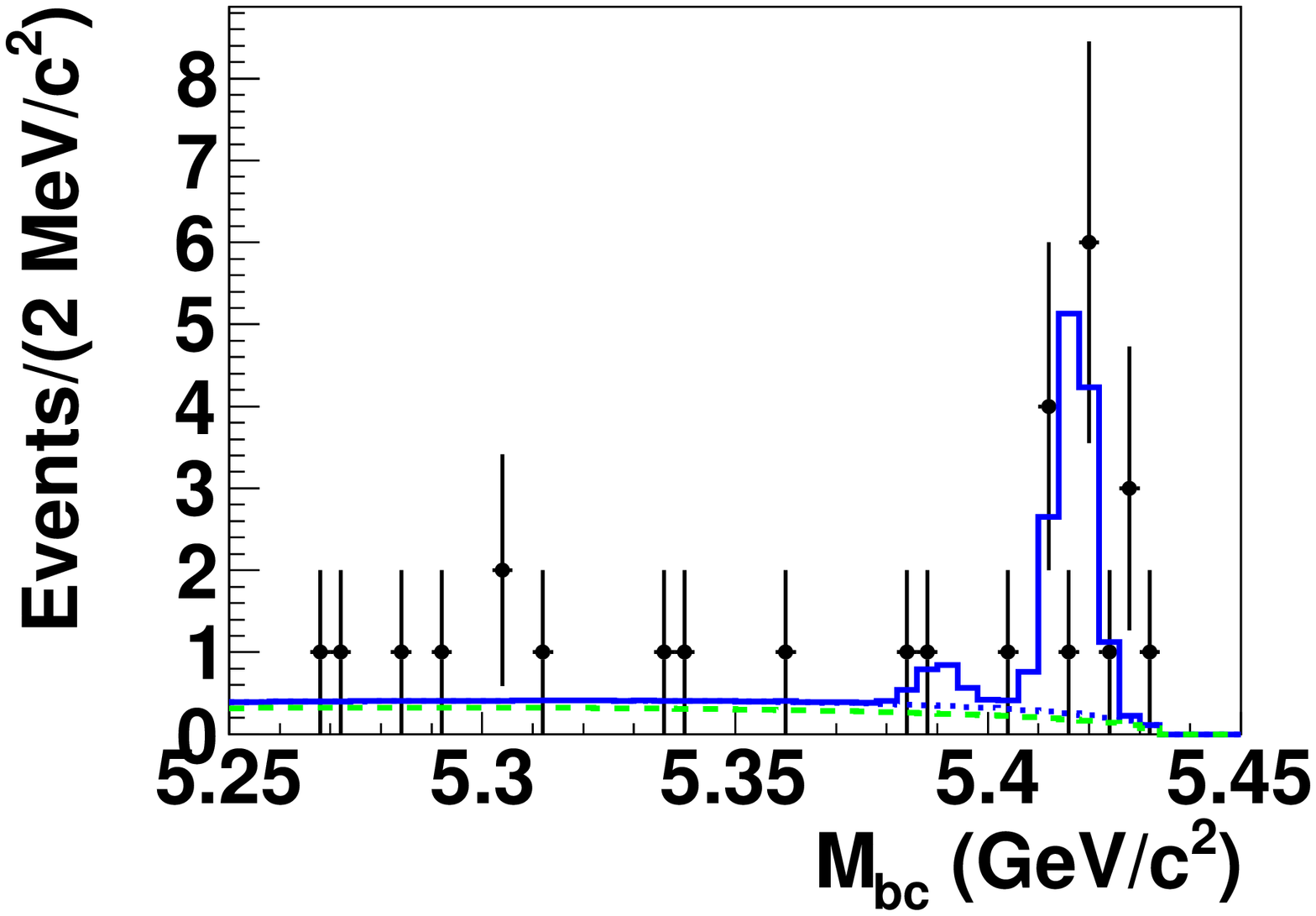}}{\epsfxsize=5.2cm\epsfbox{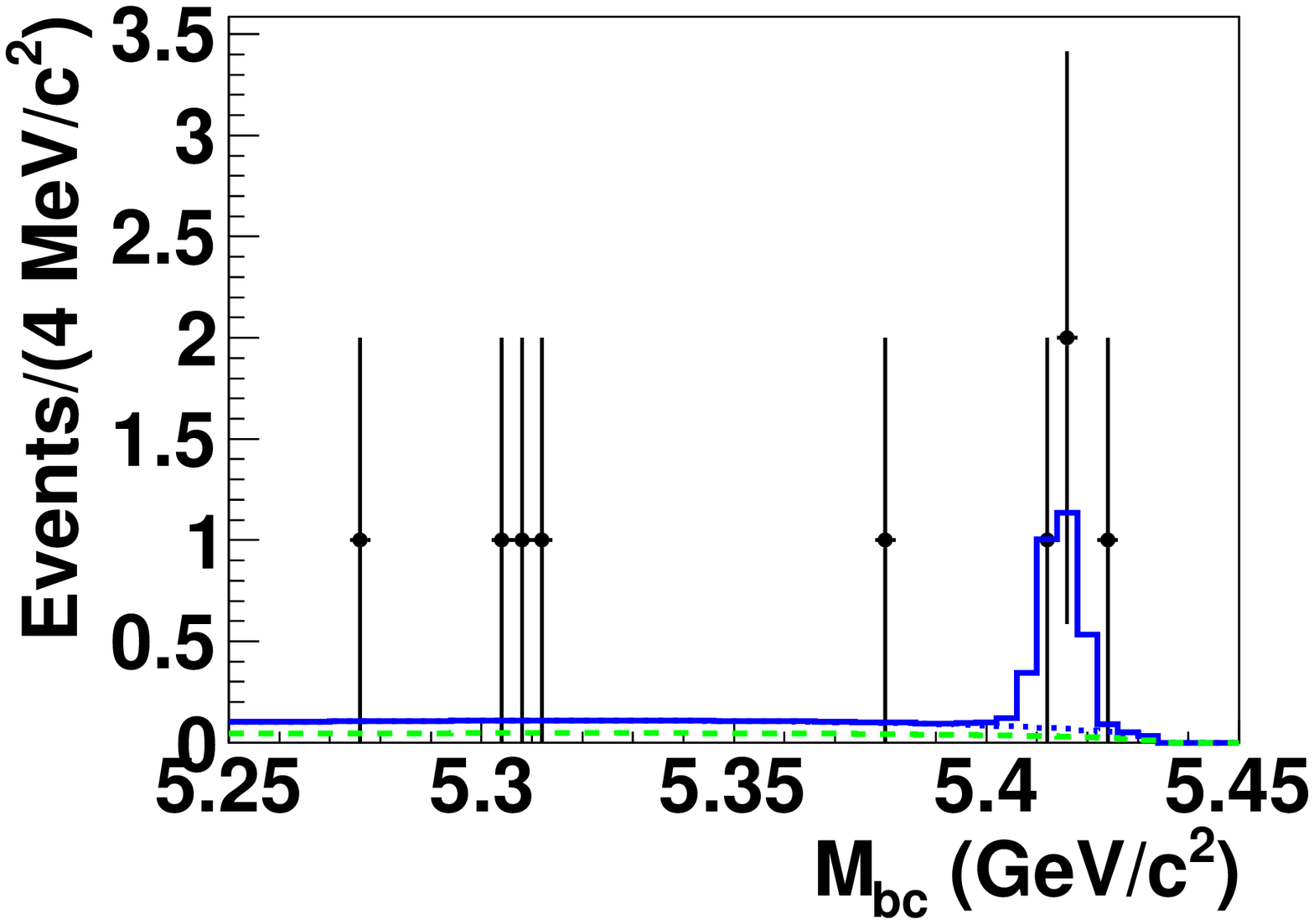}} {\epsfxsize=5.2cm\epsfbox{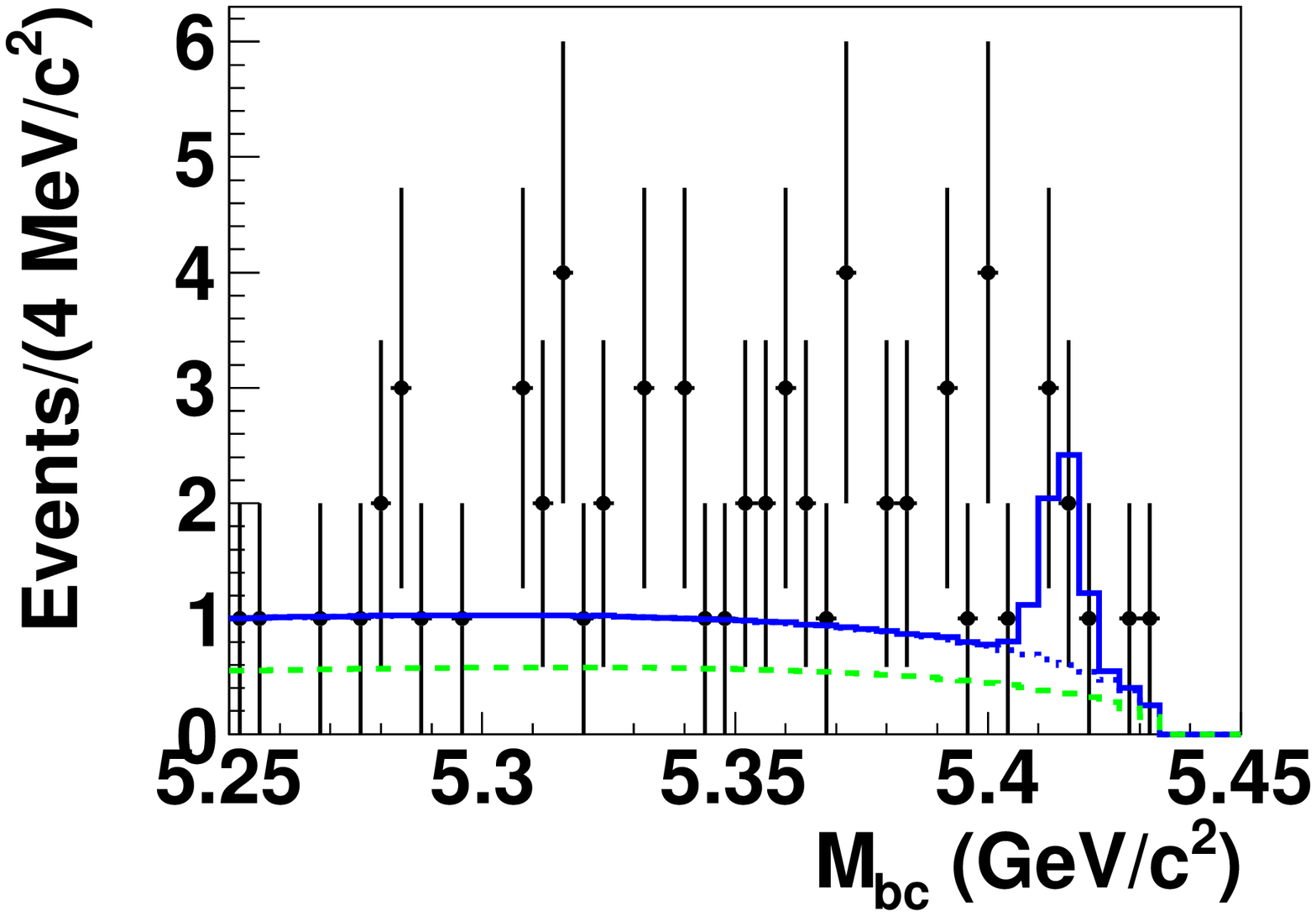}} 
\end{center} 
\caption{Projections in  $M_{\rm bc}$, based on 23.6~fb$^{-1}$ of Belle data at $\Upsilon$(10860):  (left) $B_s\to J/\psi\eta$,  (center) $B_s\to J/\psi\eta^{\prime},\eta^{\prime}\to\eta\pi\pi$, (right) $B_s\to J/\psi\eta^{\prime},\eta^{\prime}\to\rho\gamma$.
\label{psi-eta}}
\end{figure}

The decays $B_s \to J/\psi\eta^{(\prime)}$ ($CP=+1$) proceed by the same process as the $B\to J/\psi K^0$, so the branching fractions may be estimated based on the measured ${\cal B}(B_d^0\to J/\psi K^0) = 8.71\times 10^{-4}$: 
${\cal B}(B_s\to J/\psi \eta) \approx 3.5\times 10^{-4}$,
${\cal B}(B_s\to J/\psi \eta^{\prime}) \approx 4.9\times 10^{-4}$.
The decays are reconstructed in the following modes:
$J/\psi \to e^+e^-,\ \mu^+\mu^-$; $\eta \to \gamma \gamma,$ $\pi^+\pi^-\pi^0$;
$\eta^{\prime}\to \eta\pi^+\pi^-$, $\rho^0\gamma$.
The signals are extracted via a 2-dimensional fit in $\Delta E$ and $M_{bc}$.
A yield of $14.9\pm 4.1$ events ($7.3\sigma$) is observed in the channel $B_s \to J/\psi\eta$; projections in $M_{\rm bc}$ are shown in Figure~\ref{psi-eta}.
We measure 
${\cal B}(B_s\to J/\psi \eta) =(3.32\pm 0.87(stat)^{+0.32}_{-0.28}(sys)\pm 0.42(f_s))\times 10^{-4}$ and 
${\cal B}(B_s\to J/\psi \eta^{\prime}) =(3.1\pm 1.2(stat)^{+0.5}_{-0.6}(sys)\pm 0.4(f_s))\times 10^{-4}$.

We have also searched for the $CP$-eigenstate modes $B_s\to K^+K^-$, $B_s\to K^0\bar K^0$, and $B_s\to \pi^-\pi^+$, as well as the flavored mode $B_s\to K^-\pi^+$.
We find ${\cal B}(B_s\to K^+K^-) =(3.8^{+1.0}_{-0.9}(stat)\pm 0.5 \pm 0.5(f_s))\times 10^{-5}$,
${\cal B}(B_s\to K^0\bar K^0) <6.6 \times 10^{-5}$ (90\% $CL$),
${\cal B}(B_s\to K^-\pi^+) <2.6 \times 10^{-5}$(90\% $CL$),
${\cal B}(B_s\to \pi^-\pi^+)  <1.2 \times 10^{-5}$(90\% $CL$).
The findings for $K^+K^-$ and $K^0\bar K^0$ are the first absolute branching fraction and first reported limit, respectively.


\begin{figure}[b]
\begin{center}
{\epsfxsize=4.5cm\epsfbox{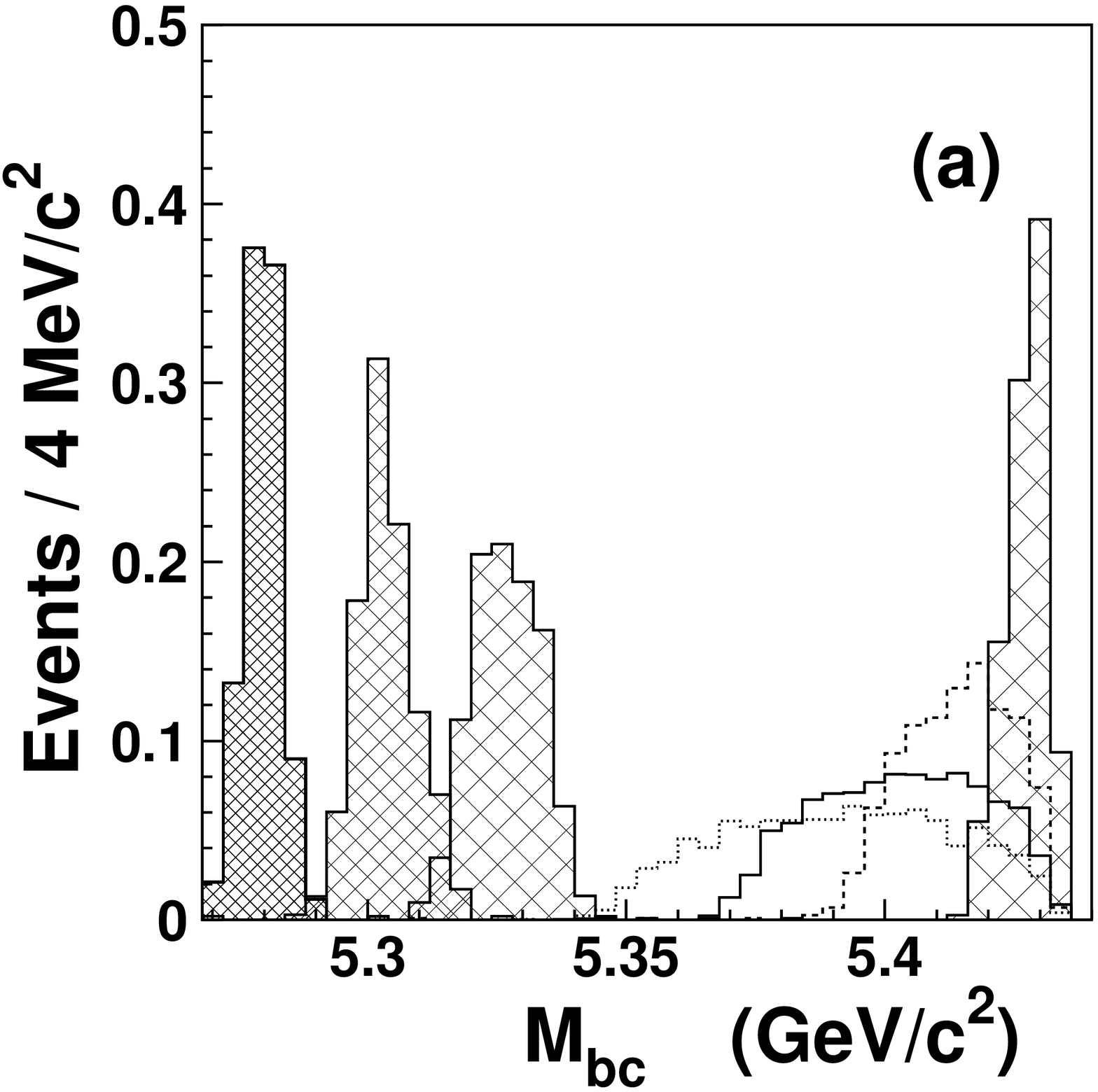}}{\epsfxsize=4.5cm\epsfbox{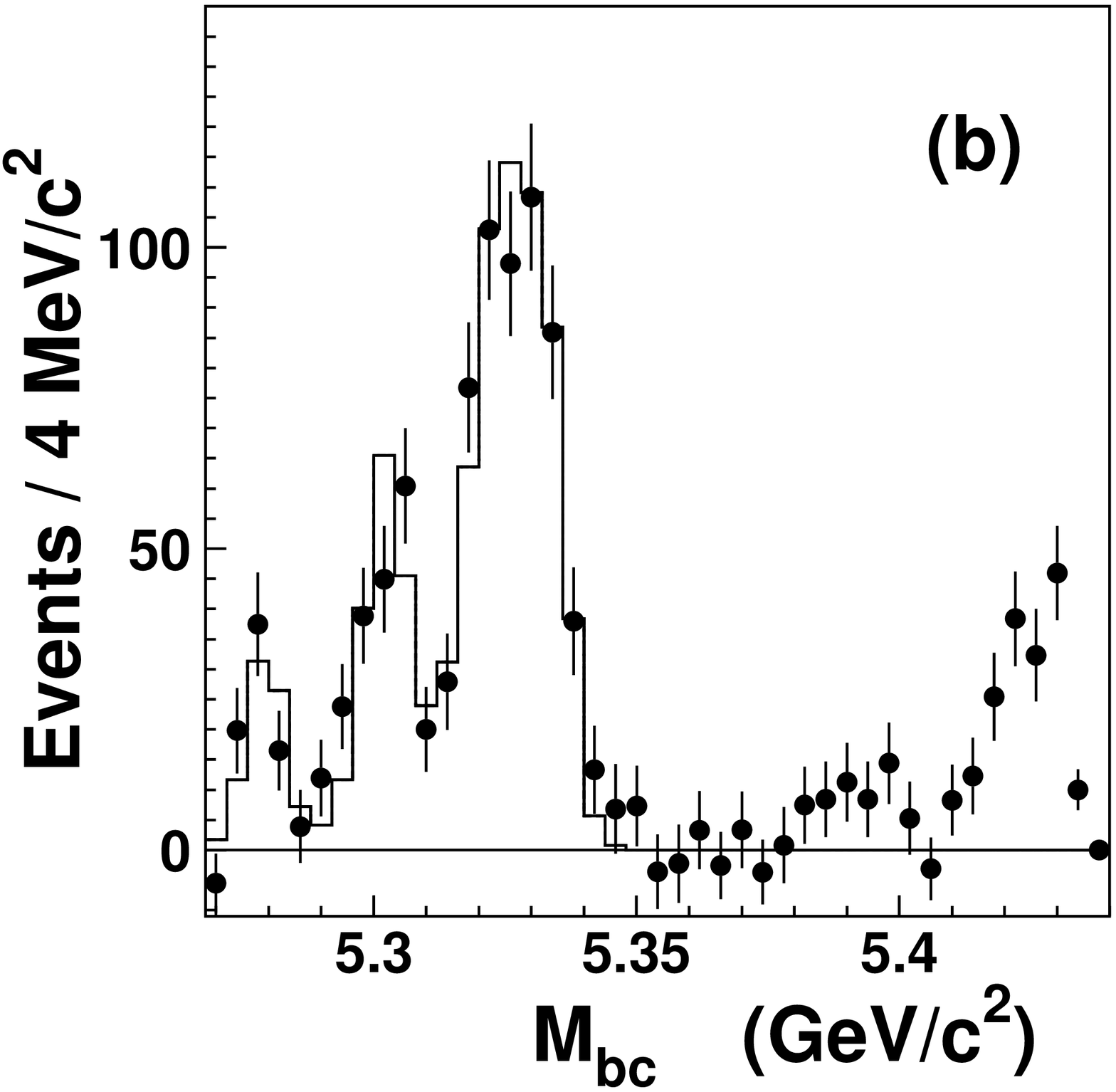}}  
\end{center} 
\caption{(a) MC simulated $M_{\rm bc}$ distributions for the 
$B^0 \to D^- \pi^+$ decay for 
$B\bar{B}$, $B\bar{B}^\ast+B^\ast\bar{B}$, $B^\ast\bar{B}^\ast$ and
$B\bar{B}\,\pi \pi$ channels
(cross-hatched histograms from left to right), and also for the three-body
channels $B\bar{B}^\ast\,\pi+B^\ast\bar{B}\,\pi$ (plain histogram),
$B\bar{B}\,\pi$ (dotted) and $B^\ast\bar{B}^\ast\,\pi$ (dashed). The distributions are normalized to unity. 
(b) $M_{\rm bc}$ distribution in data after background subtraction.
The sum of the five studied $B$ decays (points with error bars) 
and results of the
fit (histogram) used to extract the two-body channel fractions are shown.}  
\label{B-recon}
\end{figure}

\begin{figure}[ht]
\begin{center}
\psfig{figure=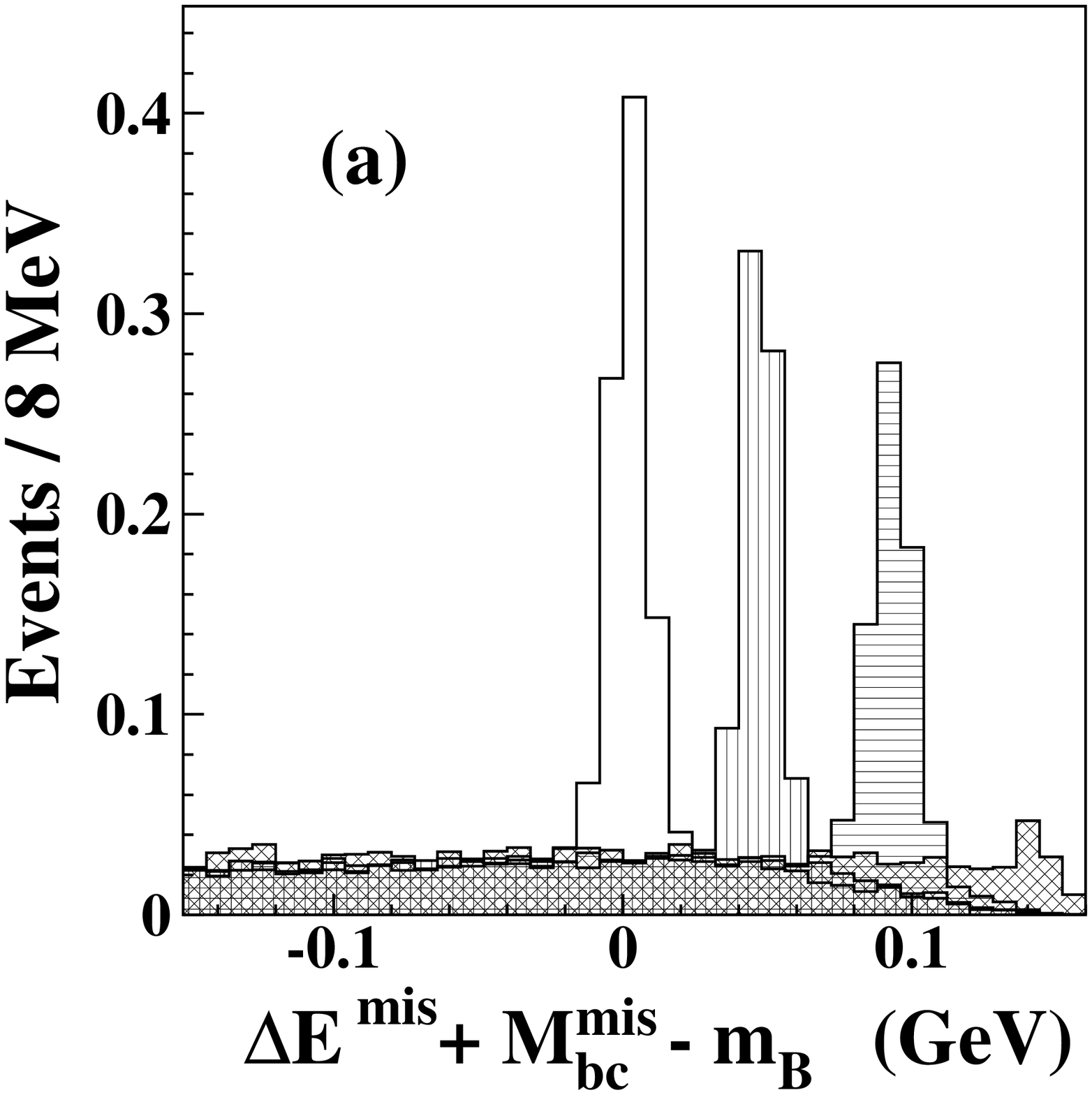,width=4.5cm}
\psfig{figure=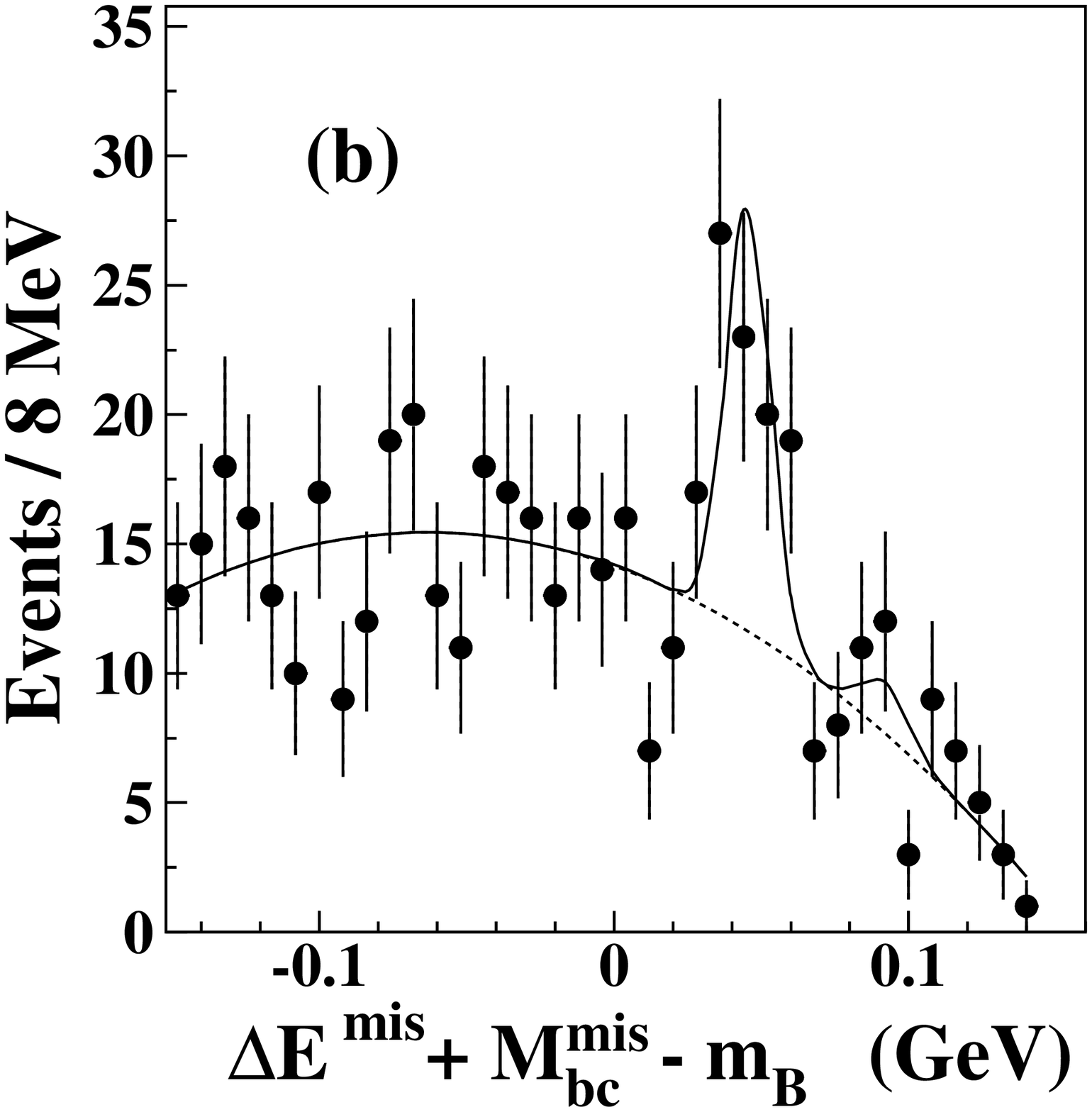,width=4.5cm}
\end{center} 
\caption{(a) The \mbox{$\Delta E^{\rm mis}+M_{\rm bc}^{\rm mis}-m_B$}
distribution normalized per reconstructed $B$ meson
for the MC simulated $B^+ \to J/\psi K^+$ decays
in the (peaks from left to right) $B\bar{B}\,\pi^+$,
$B\bar{B}^\ast\,\pi^+ +B^\ast\bar{B}\,\pi^+$,
$B^\ast\bar{B}^\ast\,\pi^+$, and $B\bar{B}\,\pi \pi$
channels.
(b) The \mbox{$\Delta E^{\rm mis}+M_{\rm bc}^{\rm mis}-m_B$} data
distribution for right-sign $B^{-/0}\,\pi^+$ combinations 
for all five studied $B$ modes. The curve shows the
result of the fit described in the text.}
\label{Bpi-frecon}
\end{figure}

While $B_s$ has been the main focus of studies at $\Upsilon$(10860), the well-tuned methods of $B$ reconstruction at the $\Upsilon$(4S) may be applied to study the more complicated assortment of $B$ events at the $\Upsilon$(10860).\cite{BBpi}
The relative rates may inform us about hadronization dynamics, and the total rate is needed to account for all $b\bar b $ events.
Neutral and charged $B$'s are reconstructed in the following modes and submodes:
$B^+\to J/\psi K^+,\ \bar D^0\pi^+$; 
$B^0\to J/\psi K^{*0},\ D^-\pi^+$;
$J/\psi \to e^+e^-,\ \mu^+\mu^-$;
$K^{*0}\to K^+\pi^-$;
$\bar D^0\to K^+\pi^-,\ K^+\pi^+\pi^-\pi^-$;
$D^-\to K^+\pi^-\pi^-$.
As with the fully reconstructed $B_s$, the signal events populate the $(\Delta E,M_{\rm bc})$ plane in clusters depending on the type of event.
Figure~\ref{B-recon}(left) shows the projections in $M_{\rm bc}$ of the distributions for the various event types.
The distribution of candidates in data, after  background subtraction, are shown in Figure~\ref{B-recon}(right).
While the distributions for events containing additional pions overlap each other, it is clear from data that their contribution is relatively small and that the majority of the rate is due to  two-body events, $B^{(*)}\bar B^{(*)}$.
It is also noted that there is an accumulation of events in the region of $B\bar B\pi\pi$, in the region of high $M_{\rm bc}$.
The fraction of $b\bar b$ events fragmenting to $B\bar B$, $B^*\bar B$, and $B^*\bar B^*$ are found to be $(5.5^{+1.0}_{-0.9}\pm 0.4)\%$, $(13.7\pm 1.3\pm 1.1)\%$, and $(37.5^{+2.1}_{-1.9}\pm 3.0)\%$, respectively.
The events where $M_{\rm bc}$ is above the two-body limit are grouped together as ``Large $M_{\rm bc}$'' and found to comprise $(17.5^{+1.8}_{-1.6}\pm 1.3)\%$.

Multibody events, in which one or more additional pions is created, may be identified by pairing reconstructed $B$'s with additional charged pions in the event and examining the {\it residual} event energy and momentum, which correspond to the opposing $B^{(*)}$ and up to one  additional pion.
From the residual 4-momentum we reconstruct $\Delta E^{\rm mis}$ and $M_{\rm bc}^{\rm mis}$.
Projections onto $M_{\rm bc}^{\rm mis}$ for various simulated event types are shown in Figure~\ref{Bpi-frecon}(left).
The corresponding distribution in data, with fit result, is shown in Figure~\ref{Bpi-frecon}(right).
The fractions of $b\bar b$ events to three-body modes $B\bar{B}\pi$, $B\bar{B}^\ast\pi$, and $B^\ast\bar{B}^\ast\pi$ are found to be  $(0.0 \pm 1.2 \pm 0.3)\%$, $(7.3\,^{+2.3}_{-2.1} \pm 0.8)\%$, and $(1.0\,^{+1.4}_{-1.3} \pm 0.4)\%$, respectively.
Paradoxically, no evidence for $B\bar B\pi\pi$ is observed, so this channel does not account for the remaining $(9.2\,^{+3.0}_{-2.8} \pm 1.0)\%$ of the ``Large $M_{\rm bc}$'' contribution observed in  $\Upsilon$(10860)$\to BX$.
The residual is quantitatively consistent with initial state radiation, $e^+e^-\to e^+e^-\gamma,\ e^+e^-\to b\bar b$, where about half the $b\bar b$ form the $\Upsilon$(4S) resonance.


In summary the Belle experiment, which was designed to  measure $CP$-asymmetry in $B$ decay, is exploring the $\Upsilon$(10860) resonance, with $\sim$120~fb$^{-1}$ data to date.
We report on 23.6~fb$^{-1}$, $\approx$1.3~million $B_s$ events.
We have made first observations of $B_s\to D_s^{*-}\pi^+$, $D_s^{(*)-}\rho^+$, $B_s\to D_s^{*+}D_s^{-}$ and $B_s\to J/\psi \eta$.
First evidence for $B_s\to D_s^{*+}D_s^{*-}$ and $B_s\to J/\psi \eta^{\prime}$ are also reported, and search for $B_s\to hh$ has yielded first limits on $B_s\to K^0\bar K^0$.
Using full reconstruction, we measure the rates for $\Upsilon$(5S)$\to BX,\ B\pi X$ and the fractions of $B_q^{(*)} \bar B_q^{(*)}$ and $B_q \bar B_q^{(*)}\pi$.
We find a residual component of 9.2\% which is consistent with being from initial state radiation.

\section*{Acknowledgments}
The author wishes to thank the organizers and staff of the 45$^{th}$ Rencontres de Moriond.
This work is supported by Department of Energy grant \# DE-FG02-84ER40153.

\end{document}